\documentclass[12pt]{article}

\usepackage{times}
\usepackage{graphicx}

\newcommand{ \lco } {La$_{2}$CuO$_{4}$ }
\newcommand{ \Lco } {La$_{2}$CuO$_{4}$, } 
\newcommand{ \lcO } {La$_{2}$CuO$_{4}$. }
 
\newcommand{ \Lbco } {(La,Ba)$_{2}$CuO$_{4}$, } 
\newcommand{ \lczmo } {La$_{2}$Cu$_{1-z}$(Zn,Mg)$_{z}$O$_{4}$ } 
\newcommand{ \Lczmo } {La$_{2}$Cu$_{1-z}$(Zn,Mg)$_{z}$O$_{4}$, } 
\newcommand{ \lczmO } {La$_{2}$Cu$_{1-z}$(Zn,Mg)$_{z}$O$_{4}$. } 
\newcommand{ \lczmod } {La$_{2}$Cu$_{1-z}$(Zn,Mg)$_{z}$O$_{4 + \delta}$ } 
 
\newcommand{ \neel } {N\'{e}el } 
 
\newcommand{ \Cel } {$^{\circ}$C } 
\newcommand{ \zcl } {$z_p$ } 
\newcommand{ \Zcl } {$z_p$, } 
 
\newcommand{ \qnlsm } {QNL$\sigma$M }
\newcommand{ \Qnlsm } {QNL$\sigma$M, } 
\newcommand{ \prl } { Phys. Rev. Lett. }
\newcommand{ \prb } { Phys. Rev. B }

\newenvironment{sciabstract}{%
\begin{quote} \bf}
{\end{quote}}

\newcounter{lastnote}
\newenvironment{scilastnote}{%
\setcounter{lastnote}{\value{enumiv}}%
\addtocounter{lastnote}{+1}%
\begin{list}%
{\arabic{lastnote}.}
{\setlength{\leftmargin}{.22in}}
{\setlength{\labelsep}{.5em}}}
{\end{list}}

\title{
Quantum Impurities in the\\
Two-Dimensional Spin One-Half\\
Heisenberg Antiferromagnet
}

\author
{
O.P. Vajk$^{1}$, P.K. Mang$^{2}$, M. Greven$^{2,3\ast}$, P.M. 
Gehring$^{4}$, J.W. Lynn$^{4}$\\
\\
\normalsize{$^{1}$ Department of Physics, Stanford University,}\\
\normalsize{Stanford, CA 94305, USA.}\\ 
\normalsize{$^{2}$ Department of Applied Physics, Stanford University,}\\
\normalsize{Stanford, CA 94305, USA.}\\ 
\normalsize{$^{3}$ Stanford Synchrotron Radiation Laboratory,}\\
\normalsize{Stanford, CA 94309, USA.}\\ 
\normalsize{$^{4}$ NIST Center for Neutron Research, National Institute}\\
\normalsize{of Standards and Technology, Gaithersburg, MD 20899, USA.}\\
\\
\normalsize{$^\ast$To whom correspondence should be addressed; E-mail:  greven@stanford.edu.}
}

\date{}

\begin{document} 

\baselineskip12pt


\maketitle 



\begin{sciabstract}
  The study of randomness in low-dimensional quantum
antiferromagnets is at the forefront of research in the field of strongly
correlated electron systems, yet there have been relatively few
experimental model systems. Complementary neutron scattering and numerical
experiments demonstrate that the spin-diluted Heisenberg antiferromagnet
\lczmo is an excellent model material for square-lattice site percolation 
in the extreme
quantum limit of spin one-half.  Measurements of the ordered moment and
spin correlations provide important quantitative information for tests of
theories for this complex quantum-impurity problem.
\end{sciabstract}

\vspace{-19cm}
\hspace{8 cm} published in {\it Science} {\bf 295}, 1691 (2002)

\newpage


        The field of low-dimensional quantum magnetism has been of
enormous interest to the condensed-matter physics community ever since the
discovery that \Lco the parent compound of the original high-temperature
superconductor \Lbco is a model two-dimen\-sional (2D) quantum (spin-1/2)
antiferromagnet. Because the superconductivity occurs in the vicinity of 
an antiferromagnetic phase in these materials, it appears likely that
antiferromagnetic fluctuations are at least partially responsible 
for their rich physics.  One of the
new frontiers in condensed-matter physics lies in the field of quantum
critical behavior, especially of "dirty" low-dimensional systems
involving quantum impurities \cite{sachdev99}.
Although there has been much progress in the experimental investigation of
quantum impurities in the simpler 1D $S=1/2$ chain 
\cite{matsuda98} and ladder
compounds \cite{azuma97}, experiments with the 2D analog have
been restricted to low impurity concentrations because of the lack of 
suitable samples \cite{hucker99,keimer92,corti95,uchinokura95}.

        We have investigated
the properties of the spin-1/2 square-lattice Heisenberg 
antiferromagnet (SLHAF) in the presence of a significant density 
$z$ of quenched, spinless quantum impurities,
up to and through the percolation threshold.
Specifically, the combined experimental and numerical results for the ordered
moment $M_{st}(z)$ and 
spin correlations $\xi(z,T)$ 
demonstrate that \lczmo is 
well described by the Hamiltonian
\begin{eqnarray}
{\cal H} = & J & \;
\sum_{\langle i, j \rangle} p_i p_j {\bf S}_i \cdot{\bf S}_{j},
\end{eqnarray}
where the sum is over nearest-neighbor (NN) sites, 
$J$ is the antiferromagnetic Cu-O-Cu superexchange, ${\bf S}_i$ is
the $S=1/2$ operator at site $i$, $p_i = 1$ on magnetic sites, and 
$p_i = 0$ on non-magnetic sites.

	In the absence of quantum fluctuations, the NN square lattice
undergoes a geometric transition with site dilution $z$ at the
percolation threshold $z_p \approx 40.725$\% 
\cite{Stauffer,newman00}. 
As indicated in Fig. 1, below this concentration there is
always one cluster of connected sites that spans the infinite
lattice.  Above \Zcl the lattice
consists entirely of finite-sized clusters.
In studies of site-diluted $S=5/2$ Heisenberg and Ising antiferromagnets, 
long-range order was found to disappear only above \zcl 
\cite{birgeneau84}.  However, in the extreme quantum limit of $S=1/2$, 
previous measurements of the N\'eel temperature $T_N$ extrapolated to
$T_N = 0$ well below \zcl \cite{hucker99,keimer92,corti95}, and the 
possible existence of a new quantum critical point at $z_{S=1/2} < z_p$
has also been raised based on theoretical \cite{yasuda99,chen00} 
and numerical \cite{miyashita92} 
considerations. Recent Monte
Carlo simulations of Eq. 1 suggest that the 
site-diluted $S=1/2$ SLHAF
remains ordered up to the percolation threshold
\cite{kato00,sandvik0110510}, 
but possibly with new, non-classical critical exponents 
\cite{kato00}. 
Our experimental data allow us to rule out the existence of a 
quantum critical point significantly below 
$z_p$, and we find that $M_{st}(z)$
is reasonably well described, up to $z \approx 30\%$, by a recent 
combined spin-wave theory and $T$-matrix approach 
\cite{chernyshev01}.

In addition to information about
the ground state properties, knowledge of the
temperature and doping dependence of the 2D instantaneous
spin-spin correlation length, $\xi(z,T)$, provides valuable
information about the dynamics of the spin degrees of freedom.
Neutron scattering studies of the spin correlations of \lco 
\cite{birgeneau99} 
and related materials
\cite{kim01,greven95},
as well as numerical results
\cite{beard98}, 
have served as important tests of
theories for Eq. 1 in the absence of quantum impurities
\cite{chakravarty89,hasenfratz91}. 
Remarkably, even though the quenched
disorder leads to the loss of Lorentz invariance, a quantum non-linear
sigma model approach developed for Eq. 1 in the absence of disorder
is found to give an excellent effective description of $\xi(z,T)$
up to at least $z=35\%$.
Our experimental and numerical results can 
help guide the development of an underlying theory for the
randomly diluted system.

Several experimental difficulties have previously prevented quantitative
experimental studies of randomly diluted \lcO
First, the highest reported concentration
of non-magnetic ions is 25\% 
\cite{hucker99},
still well below $z_p$.
Second, single crystal results have been limited even further,
to $z \approx 15$\%
\cite{keimer92,uchinokura95}.
Third, the excess oxygen
typically found in as-grown samples introduces holes into the copper-oxygen
sheets, which frustrate the antiferromagnetism and quickly destroy
magnetic order 
\cite{aharony88}.
Differing values for $T_N (z)$ 
indicate that this problem has not been fully resolved
\cite{hucker99,keimer92,corti95,uchinokura95}.

By jointly substituting Zn and Mg on the Cu site, we were able to 
grow\linebreak
\lczmod crystals by the traveling-solvent floating-zone method.
The Zn content was approximately $10$\%
whereas the Mg content varied.
Typical single grain sections were 40 mm long and 4 mm in diameter.
Both Zn$^{2+}$ and Mg$^{2+}$ are non-magnetic,
effectively removing a
magnetic site without introducing charge carriers.
Zn$^{2+}$ has a larger ionic radius than Cu$^{2+}$, and
Mg$^{2+}$ has a smaller ionic radius than Cu$^{2+}$.
Compositional analysis was carried out
by electron probe microanalysis on end sections of the crystals.
The samples were carefully reduced at $T = 900$ to $950$ \Cel in Ar flow.
Several crystals used for neutron scattering had a small Cu/Zn/Mg
concentration gradient, typically 1 to 2\% dilution, along the 
full length of the
crystal. Mosaic widths were very good, 15' full width at
half maximum (FWHM) or less.
Small, very homogeneous sections
a few cubic millimeters in size were cut from the
larger crystals for magnetometry.
Polycrystalline samples, with concentrations assumed to be equal to
their nominal values, 
yielded similar magnetometry results.

Elastic scattering at the $(1,0,0)$ Bragg peak (orthorhombic notation),
shown in Fig. 2A, was used to determine $T_N$ and $M_{st}(z)$.
We find long-range \neel order
even at $z = 39$\%, where $T_N$ is only 8 K.
Figure 2B shows our result for $T_N (z)$.
Below $z\approx20$\%, the data agree well with the linear behavior
found in previous experimental 
\cite{hucker99,keimer92,corti95} 
and recent theoretical 
\cite{chernyshev01} 
work. However, we find that
$T_N$ falls off more gradually at higher concentrations. Quantum
fluctuations for $S=1/2$ apparently are not strong enough to
noticeably shift the critical point: $z_{S=1/2} = z_p$, within the
uncertainty of our experiment.

In order to probe the ground state properties of \lczmo we have
extracted $M_{st}(z)$ from the low-temperature (1,0,0) intensity.
To facilitate normalization between
samples, we measured the intensity of a standard phonon for each
sample to determine the illuminated volume; 
normalizing by sample mass gives the same result within the errors.  
The resulting normalized moment per Cu atom is plotted in Fig. 2C.
Our data are consistent with previous nuclear quadrupole resonance(NQR) 
measurements below $z=15\%$ \cite{corti95}. 
Numerical results for Eq. 1 deviate somewhat
in their magnitude, but also suggest that $z_{S=1/2} = z_p$
\cite{kato00,sandvik0110510}. 
Recent theory 
\cite{chernyshev01}, 
which uses a spin-wave theory and $T$-matrix approach and is expected to 
be valid at low and intermediate concentrations, gives a good
description of our data up to $z\approx30\%$. The 
classical ($S \rightarrow \infty$) behavior is significantly different,
because the reduction of the moment is attributed
only to Cu sites being disconnected from the infinite cluster.
It should be noted that already $M_{st}(0)$ is reduced from the
classical value $M_{st,cl}(0) = 1/2 $ by about 40\% because of
quantum fluctuations 
\cite{beard98,yamada87}. 
These fluctuations
evidently increase with dilution, further weakening the ordered
moment, but not sufficiently to disorder the system below $z_p$.
The experimental data are well described by a power law with
$z_{S=1/2} = z_p$ and an effective exponent $\beta_{eff} = 0.45(3).$
In their finite-size scaling analysis, Kato et al.
\cite{kato00} 
extracted spin-dependent critical exponents, with
$\beta = 0.46$ for $S=1/2$, which is substantially different from
the classical value $\beta_{cl} = 5/36$
\cite{Stauffer}. 
However, it has been argued that
these are not the true critical exponents, which should equal the
classical values
\cite{sandvik0110510}.

We also performed a systematic study of the instantaneous 2D
spin-spin correlation length $\xi(z,T)$ in the paramagnetic phase
of \lczmO  The equal-time structure factor was
measured in two-axis, energy-integrating mode 
\cite{birgeneau99}.
Figure 3A shows representative data for $z=19\%$ at $T=200$ K.
The scattering broadens as $\xi$
decreases, both with increasing temperature and increasing
dilution.  Correlation lengths were
obtained from fits to a Lorentzian, $\sim 1/(1 + q_{2D}^2 \xi^2)$,
where $q_{2D}$ is the 2D momentum transfer component in the
CuO$_2$ sheets relative to the zone center at $H=1$, convoluted
with the instrumental resolution. The extracted lengths are plotted
versus $J/T$ in
Fig. 3B. The data shown were cut off above $T_N$ by 1 SD
($\approx 4$ K), as obtained from fits of the
order parameter assuming a Gaussian distribution in $T_N$
(Fig. 2A). Temperature is scaled by $J = 135$ meV, the
antiferromagnetic superexchange energy of the pure system
\cite{birgeneau99}.
The data reveal that doping
significantly decreases the rate at which correlations grow as the
system is cooled. Specifically, at high concentrations $\xi(z,T)$
crosses over from exponential to power-law behavior. We note that
the $z=40(2)$\% and 43(2)\% samples do not exhibit N\'eel order at
1.4 K ($J/T \approx 1100$),
and the spin correlations appear to approach a constant
zero-temperature value, as expected for $z>z_p$.

In order to test the degree to which the experimental system is
described by Eq. 1, we have performed quantum Monte Carlo (QMC)
simulations to calculate $\xi(z,T)$. We used the loop-cluster algorithm,
which is suited to study the randomly diluted Heisenberg model
\cite{greven98}.
Lattice sizes 10 to 20
times larger than the correlation length were used in order to
avoid finite-size effects. We were able to perform our
calculations on very large lattices of up to $1700 \times 1700$ sites
and to temperatures as low as $T = J/100$. 
Previous numerical work involved system sizes as large as $20 \times 20$
\cite{miyashita92,manousakis92}
and reached $T = J/2$ 
\cite{manousakis92} 
and $T = J/20$ 
\cite{miyashita92}.
Between 5 and 200
configurations were averaged for each temperature and
concentration, and we chose $10^4$ to $10^5$ equilibrations and
$10^4$ to $10^5$ measurements per configuration. 
The QMC results (Fig. 3B) extend to higher temperatures and thus complement the 
experiment, covering a combined three orders of magnitude in temperature.  
We emphasize that this comparison contains no
adjustable parameters, because $J$ is known rather well for the
experimental system. We find excellent quantitative agreement up
to the percolation threshold.

The ground state of the pure SLHAF is ordered, but quantum
fluctuations renormalize the spin-wave velocity, $c =
2 \sqrt{2} S Z_c J a$, and spin-stiffness, $\rho_s = S^2 Z_{\rho}J$,
from their classical values (we use
units in which $g\mu_B = k_B = \hbar = 1$). For $S=1/2$ and $z=0$, the quantum
renormalization factors $Z_c$ and $Z_{\rho}$ are known from
various theoretical and numerical studies. Using these quantities,
$\xi (z=0,T)$
is quantitatively given by
\cite{chakravarty89,hasenfratz91}
\begin{equation}
\frac{\xi}{a} = \frac{e}{8} \frac{c/a}{2 \pi \rho_s} e^{2 \pi \rho_s/T}
\left[1 - \frac{1}{2} \left( \frac{T}{2 \pi \rho_s} \right)
+ {\cal O} \left(\frac{T}{2 \pi \rho_s} \right)^2 \right].
\end{equation}
Even though Eq. 2 is strictly valid only at asymptotically low
temperatures 
\cite{beard98,chakravarty89,hasenfratz91}, 
it agrees remarkably well with experiment 
\cite{birgeneau99,kim01,greven95} 
and numerics 
\cite{beard98} 
for $S=1/2$ in the range $2 < \xi/a < 200$
shown in Fig. 3B. The derivation of Eq. 2
involves a mapping of the discrete Heisenberg Hamiltonian Eq. 1
to a quantum non-linear sigma model (QNL$\sigma$M),
and it is based on the
assumptions of an ordered ground state and of translational
invariance. Random dilution breaks translational invariance of the 
SLHAF and leads to defect rods in the time-like direction of the 
effective QNL$\sigma$M. The latter implies the loss of Lorentz 
invariance, and a
QNL$\sigma$M description 
may no longer remain valid 
\cite{chakravarty89,chernyshev01}. 
Nevertheless, it is
valuable to test the extent to which this description may hold.
Because there exist no accurate predictions for $c(z)$ and
$\rho_s(z)$, such a test requires us to treat these 
zero-temperature quantities as fit parameters.
A modified form of Eq. 2,
\begin{equation}
\frac{\xi}{a} = \frac{e}{8} \frac{c/a}{2 \pi \rho_s}
\frac{e^{2 \pi \rho_s/T}}
{1 + (4 \pi \rho_s/T)^{-\nu_T}}
\end{equation}
with $\nu_T = 1$, has been suggested for disorder-free systems approaching 
a quantum critical point 
\cite{castroneto96}, 
where $\rho_s =0$ and
$\xi \sim 1/T$ 
\cite{chakravarty89}. 
Even though $z_{S=1/2} = z_p$ may not be a quantum
critical point, random dilution reduces the spin stiffness
and may be viewed as bringing the
system closer to such a point in an extended parameter space. 
In a classical picture, $z=z_p$ is a (geometric and thermal)
multi-critical point, and $\rho_s = 0$ as well as
power-law behavior are expected. The thermal correlation length exponent 
$\nu_T = 0.90(5)$ was previously determined for the $S=5/2$ system
Rb$_2$(Mn,Mg)F$_4$ 
\cite{birgeneau84}.
Furthermore, for $z>z_p$ it was found that the data were well
described by the form
\begin{equation}
\frac{1}{\xi(z,T)} = \frac{1}{\xi_0(z)} + \frac{1}{\xi_T(T)}
\end{equation}
where $\xi_0$ is a zero-temperature length, possibly reduced from
the geometric length of the percolation problem by small quantum
fluctuations, and $\xi_T \sim T^{-\nu_T}$. 

We find that Eq. 2 gives a very good description of our data, 
especially at lower concentrations, and for $\xi/a > 8$ to $10$ 
at $z =31\%$ and 35\%.
The modified form Eq. 3, however, 
even captures the high-temperature
power-law behavior at the higher concentrations.
Results of fits to Eq. 3
of our QMC data for $z<z_p$ are shown in Fig. 3B
and summarized in Table I. 
Fits to Eq. 2 result in large uncertainties in 
the value of the spin-wave velocity, but $\rho_s (z)$ and $c(z)$ 
extracted using these two forms agree within the errors.
For the pure system, a fit of numerical data 
\cite{beard98} 
below $\xi/a = 200$
yields $2 \pi \rho_s (0) = 1.18(1)J$
and $c(0) = 1.33(3)Ja$, about
4\% higher and 20\% lower, respectively, than the most accurate estimate 
\cite{beard98}. 
A recent combined \Qnlsm and  percolation theory approach
\cite{chen00} resulted in 
\begin{equation}
\frac{\rho_s(z)}{\rho_s(0)} = A(z) \left[1 - 
\frac{\bar{g}(0)}{P_{\infty}(z)} \right]
\frac{1}{1-{\bar{g}(0)}},
\end{equation}
where $\bar{g}(0) = 0.685$ is the coupling constant 
corresponding to the $S=1/2$ SLHAF at $z=0$ 
\cite{chakravarty89},
and $A(z)$ and $P_{\infty}(z)$ are the bond dilution factor
and the probability of finding a spin in the infinite cluster. The
latter two quantities are well described
up to $z \approx 37\%$ and $z \approx 30\%$, respectively, by
$A(z) \approx 1-\pi z+\pi z^2/2$ 
and $P_{\infty}(z) \approx 1 - z$
\cite{Stauffer,harris77}.
Equation 5 incorrectly predicts $\rho_s(z\approx30\%) = 0$, and hence
a quantum critical point well below the percolation threshold
\cite{chen00}.
To our surprise, we find that substituting $1+z$ for $1/P_{\infty}(z)$
{\it quantitatively} describes $\rho_s(z)/\rho(0)$ even at $z=35\%$, as
shown in Table 1.  This substitution is correct at low concentrations
and prevents the second term in Eq. 5 from going to zero below
$z_p$. 
We note that Eq. 5 is a one-loop renormalization-group result,
and higher-order terms can be expected to improve agreement with our
observations.
The expression for the spin-wave velocity that corresponds to Eq. 5
\cite{chen00},
$c(z)/c(0) = A(z) (1 + z/2)$,
decreases monotonically with dilution, and hence does not describe
the behavior found from our fits (see Table 1).

For $z>z_p$, the joint experimental ($z=40(2)\%$ and $z=43(2)\%$) and
numerical ($z=41\%$ and $z=46\%$) data are fit to Eq. 4. Better
results are obtained if $1/\xi_0$ is allowed to be non-zero:
$\xi_0(z \approx 41\%)/a = 50(20)$ and $\nu_T (z\approx 41\%) = 0.72(7)$. A
value of $\nu_T \approx 0.7$, together with $z_{S=1/2} = z_p$, 
is also obtained in a scaling analysis
of our numerical data.  
We note that this value lies
below $\nu_T = 0.90(5)$ for the $S=5/2$ system Rb$_2$(Mn,Mg)F$_4$
\cite{birgeneau84}. 
Fits to the data at the highest
concentration give $\nu_T=0.70(7)$ and $\xi_0/a = 7(1)$.

The percolation threshold $z_p \approx 40.7\%$ is for the NN square 
lattice, and a non-zero frustrating next-NN (NNN) exchange 
could, in principle, shift $z_p$ and lead to low-temperature spin-glass 
physics.  The NNN exchange in \lco is expected to be frustrating and 
may be as large as $8\% J$
\cite{kim01}, 
which would lead to a small reduction of the ordered moment, 
but should not shift the critical point 
\cite{birgeneau84}.
The dominant further-neighbor interaction may actually be a ring exchange
among the four Cu sites on a square 
\cite{coldea01}, 
which cannot extend connectivity beyond
the NN percolation threshold. In any case, further-neighbor
interactions do not seem to noticeably affect the nature of the 
transition in \Lczmo, $z_{S=1/2} = z_p$ 
within the uncertainty of our experiment,
and we find no evidence of spin-glass behavior.

The correction terms to Eq. 1
can be expected to change as a result of the observed evolution of the
low-temperature structure. Above $z\approx 25$\%, N\'eel order
occurs in a low-temperature tetragonal phase.
Nevertheless, we find that $T_N(z)$ evolves smoothly. 
Interestingly, our QMC data for Eq. 1 indicate that,
as for undoped \lco \cite{birgeneau99}, 
$T_N(z)$ corresponds very well to the temperature at which
$\xi_{2D}/a = 100$,
as shown in Fig. 2B.
This suggests that any changes in the correction terms
in the full spin Hamiltonian must be very
subtle. 

Assuming that quantum fluctuations do not
alter the classical $T=0$ correlation length exponent, $\nu_0 = \nu_{cl} = 4/3$
\cite{Stauffer}, 
our result of $\nu_T \approx 0.7$ suggests
a crossover exponent $\Phi = \nu_0/\nu_T \approx 1.9$, which is
larger than the values $\Phi=1.43$ to $1.7$ predicted theoretically
\cite{stanley76,coniglio81}. 
On the other hand, if $z=z_p$ is a quantum critical point,
it would suggest a dynamical
critical exponent of $z = 1/\nu_T \approx 1.4$.
Series expansion and numerical approaches
for the $S=1/2$ bond percolation problem,
which is closely related to the site percolation problem of the present study,
predict $z = 1.7$ to $2$ 
\cite{wan91}, 
whereas recent QMC work for
the site-diluted $S=1/2$ SLHAF arrives at $z \approx 2.5$
\cite{kato00}. 
Our value is lower than these results, but also 
lies significantly above $z = 1$ for the Lorentz-invariant \qnlsm
\cite{chakravarty89}, 
implying that the \qnlsm fixed point is
unstable to randomness.

Recent theoretical work for the randomly diluted $S=1/2$ SLHAF 
has led to some exact results in the dilute limit of quantum 
impurities 
\cite{sachdev99}, 
and to interesting new predictions
at higher concentrations 
\cite{chen00,chernyshev01}.
It has been argued that the presence of impurities should lead
to localized spin excitations and 
to the breakdown of the classical hydrodynamic description of
excitations in terms of spin waves above a characteristic length scale
\cite{chernyshev01}.
In this picture, the spin-wave velocity $c(z)$ is not a well-defined
quantity, but there is no instability toward a disordered phase.
Unlike dynamic observables, static properties such as the 
staggered magnetization and $T_N$ remain well
defined throughout the ordered phase.
Because the spin-stiffness also remains well defined in this
theory, one continues to expect the correlation length to 
have the low-temperature form  $\xi \sim e^{2\pi\rho_s/T}$ as 
for the pure system.
We have been able to determine $\xi(z,T)$ over a very wide range of 
impurity concentrations
and temperature, and arrive at the unexpected conclusion that 
Eq. 2, and especially the heuristic crossover form Eq. 3, provide
an excellent description of our data.


\begin{scilastnote}
\item 
O.P.V. and M.G. thank A. Aharony, A. H. Castro Neto, 
A. L. Chernyshev, A. W. Sandvik, and E. F. Shender for helpful 
discussions. The work at Stanford was supported by the U.S. Department of 
Energy under contract nos. DE-FG03-99ER45773 and DE-AC03-76SF00515, by NSF 
CAREER Award no. DMR9400372, and by the by the A.P. Sloan Foundation.
\end{scilastnote}

\clearpage

\begin{table}
    \caption[~]{Spin stiffness and spin-wave velocity extracted from
        fits of numerical data
        to Eq. 3.  For $z < 35\%$, $\nu_T$ was fixed to be 1.
        At $z = 35\%$, a better fit was obtained using
        $\nu_T = 0.88(1)$.  Equation 5 is modified by substitution
        of $1 + z$ for $1/P_{\infty}$, as discussed in the text.  Theory
        for $c(z)$ is given by $c(z)/c(0) = A(z)(1+z/2)$, as described in
        the text.
        The values for $2 \pi \rho_s(0)$ and $c(0)$ are from
        \cite{beard98}.\\
} 
    \label{fits_table}
    \begin{tabular}{rllll}
    $z$ & $2\pi\rho_s(z)/J$ & Modified Eq. 5 & $c(z)/J/a$ & $\rm Theory$\\ 
    0 & 1.18(1)  & 1.13 & 1.33(3)  & 1.66 \\
    0.08 & 0.79(1)  & 0.71 & 1.35(8)  & 1.31 \\
    0.20 & 0.31(1)  & 0.28 & 1.43(9)  & 0.79 \\
    0.31 & 0.063(1) & 0.065 & 1.03(2)  & 0.34 \\
    0.35 & 0.026(1) & 0.025 & 0.69(1)  & 0.18 \\
\end{tabular}
\end{table}

\begin{figure}
\centering
\includegraphics[width=12cm]{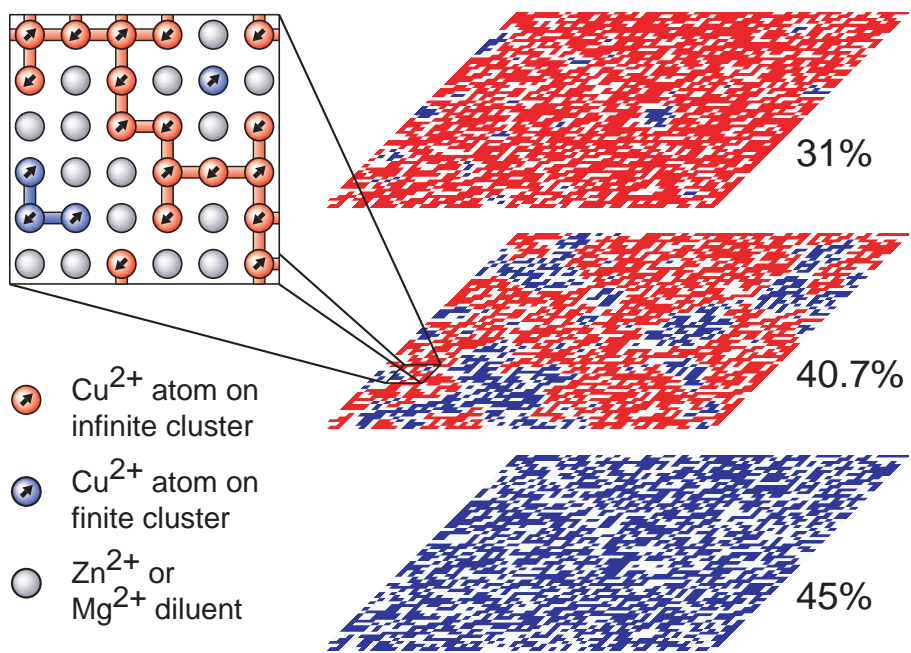}
\caption{\label{percolation} 
Schematic of finite-sized sections of the infinite square lattice with
random site dilution levels well below (31\%), just below (40.7\%), and
above (45\%) the percolation threshold $z_p \approx 40.725$\%
\cite{newman00}.  Sites on the infinite cluster are shown in
red, sites on finite disconnected clusters in blue, and diluents in white.  
The inset is a close-up view for $z=40.7$\%, showing the role magnetic Cu
and non-magnetic Zn/Mg ions play in the experimental
system. 
}
\end{figure}

\begin{figure} 
\centering 
\includegraphics[width=6cm]{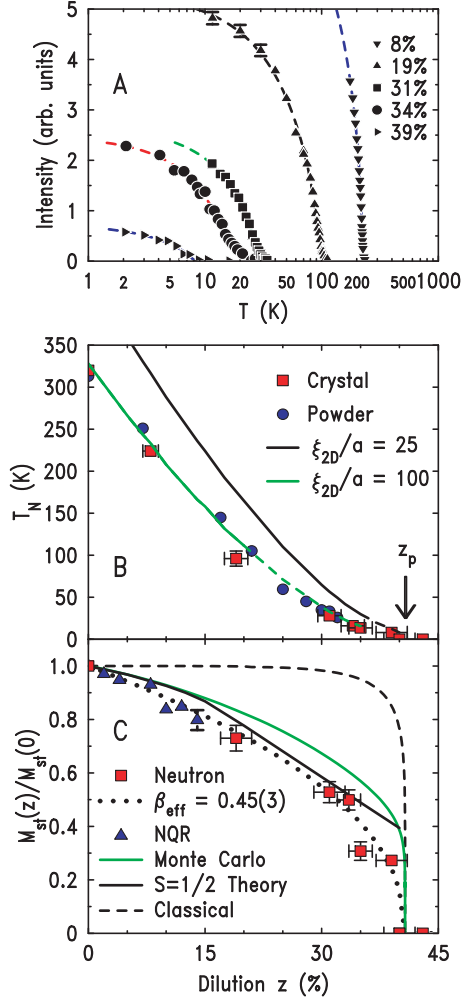}
\caption{\label{3panel} 
{\bf (A)} (1,0,0) magnetic peak
         intensity from neutron diffraction (orthorhombic notation).
         A temperature-independent background has
         been subtracted.  Lines represent fits for the magnetic order
         parameter squared, $\sim (T_N - T)^{2\beta}$ with
         $\beta \approx 0.30$, assuming a Gaussian distribution of
         $T_N$ (typically $\approx 4$ K) to describe the rounding due
         to the small inhomogeneities present in the large samples used.
{\bf (B)} $T_N$ from neutron diffraction (crystal) and
         magnetometry (crystal and powder sample).
         Above $z \approx 20\%$, $T_N (z)$ deviates from an extrapolated
         line, approaching zero at the percolation threshold $z_p$.
         Lines correspond to constant correlation lengths
         $\xi_{2D}/a$ = 25 and 100 from Monte Carlo simulations of the
         randomly diluted $S=1/2$ NN SLHAF; dashed regions are
         extrapolated from higher temperature.
{\bf (C)}  Staggered moment per Cu atom, normalized by the
         value for the pure system. The neutron data are consistent
         with previous NQR results \cite{corti95}.
         The data are well described by a power law with $z_{S=1/2} = z_p$
         and exponent $\beta_{eff} = 0.45(3)$. Also shown is recent QMC
         \cite{sandvik0110510} and theory \cite{chernyshev01} for $S=1/2$,
         as well as the classical ($S \rightarrow \infty$) result
         \cite{sandvik0110510}. 
}
\end{figure}

\begin{figure}
\centering
\includegraphics[width=12cm]{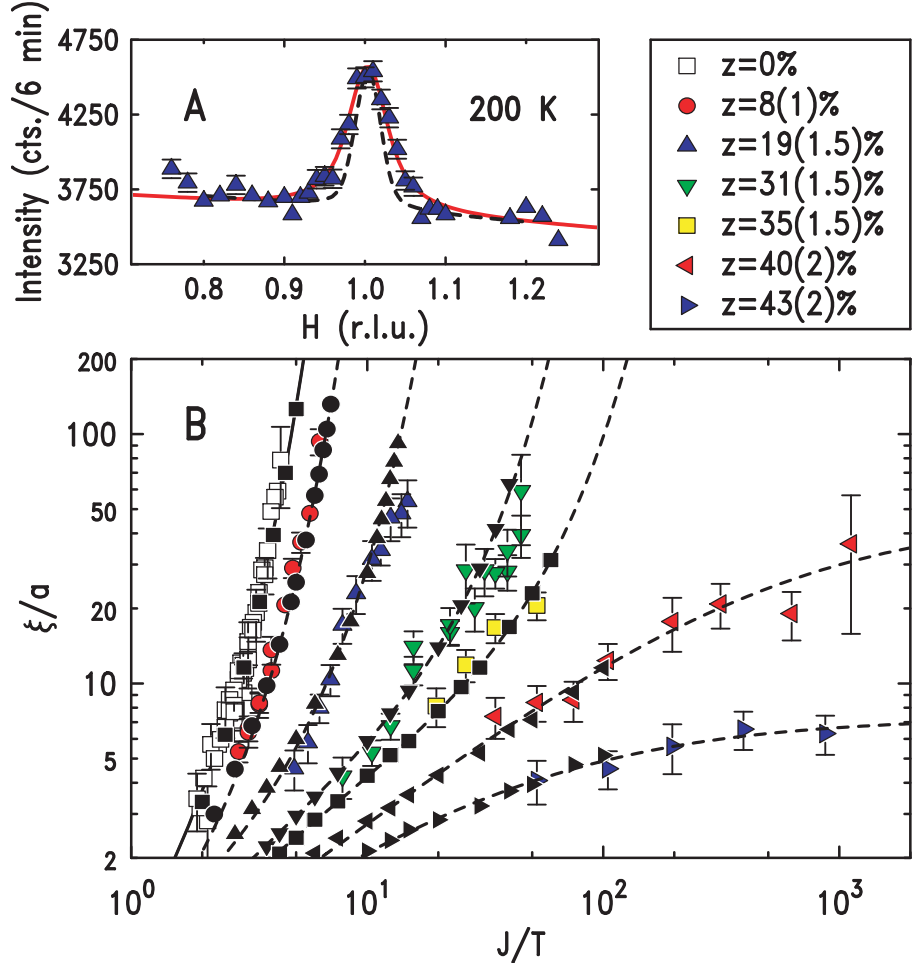}
\caption{\label{xi} 
    {\bf (A)} Representative equal-time structure factor data.
    The solid red line represents a fit to the data, as
    discussed in the text, and the dashed black line indicates
    the instrumental resolution.  This measurement
    was carried out in two-axis mode with 30.5 meV
    incident neutron energy and
    horizontal collimations of 40'-27.5'-sample-23.7'.
    {\bf (B)} Spin-spin correlation lengths in units of the 
    lattice constant. Colored symbols represent
    results from neutron scattering measurements of
    La$_{2}$Cu$_{1-z}$(Zn,Mg)$_{z}$O$_{4}$;
    black symbols represent Monte Carlo data
    for $z=8,20,31,35,41$, and 46\%.  No adjustable
    parameters were used in the comparison.
    Experimental and numerical results for $z=0$
    are from \cite{birgeneau99,beard98} and the
    solid line is Eq. 2.
    Dashed lines are fits to
    Eqs. 3 and 4, as discussed in the text.  
}
\end{figure}

\end{document}